\documentclass[twocolumn,preprintnumbers,amsmath,amssymb]{revtex4}
\usepackage{tabularx,graphicx}

\usepackage{color}
\usepackage{hyperref}
\hypersetup{
    colorlinks=true,
    linkcolor=blue,
    filecolor=blue,      
    urlcolor=blue,
}

\usepackage{color}

\usepackage{ulem}   

\begin{document}

\newcommand{\beq}{\begin{equation}}
\newcommand{\eeq}{\end{equation}}
\newcommand{\beqn}{\begin{eqnarray}}
\newcommand{\eeqn}{\end{eqnarray}}
\newcommand{\bmath}{\begin{subequations}}
\newcommand{\emath}{\end{subequations}}
\newcommand{\bra}[1]{\langle #1|}
\newcommand{\ket}[1]{|#1\rangle}

\title{On the interpretation of  flux trapping experiments in hydrides}

\author{J. E. Hirsch$^{a}$  and F. Marsiglio$^{b}$ }
\address{$^{a}$Department of Physics, University of California, San Diego,
La Jolla, CA 92093-0319\\
$^{b}$Department of Physics, University of Alberta, Edmonton,
Alberta, Canada T6G 2E1}
 
  \begin{abstract} 
  In Ref. \cite{etrappedp},  Minkov et al reported measurements of the magnetic moment that remains after a magnetic field is turned on and then
 turned off for hydride materials under high pressure in a diamond anvil cell.
In Refs.  \cite{e2021p,correction}, Minkov et al reported magnetization measurements on the same samples as a function
of applied magnetic field. Here we argue  that the latter indicate that the signal measured in the former does not 
provide evidence for superconductivity in these samples. Instead, the measured signal likely originates in
ferromagnetism of either the sample or the background.

  \end{abstract}
  \maketitle 
 
          \begin{figure} [t]
 \resizebox{8.5cm}{!}{\includegraphics[width=6cm]{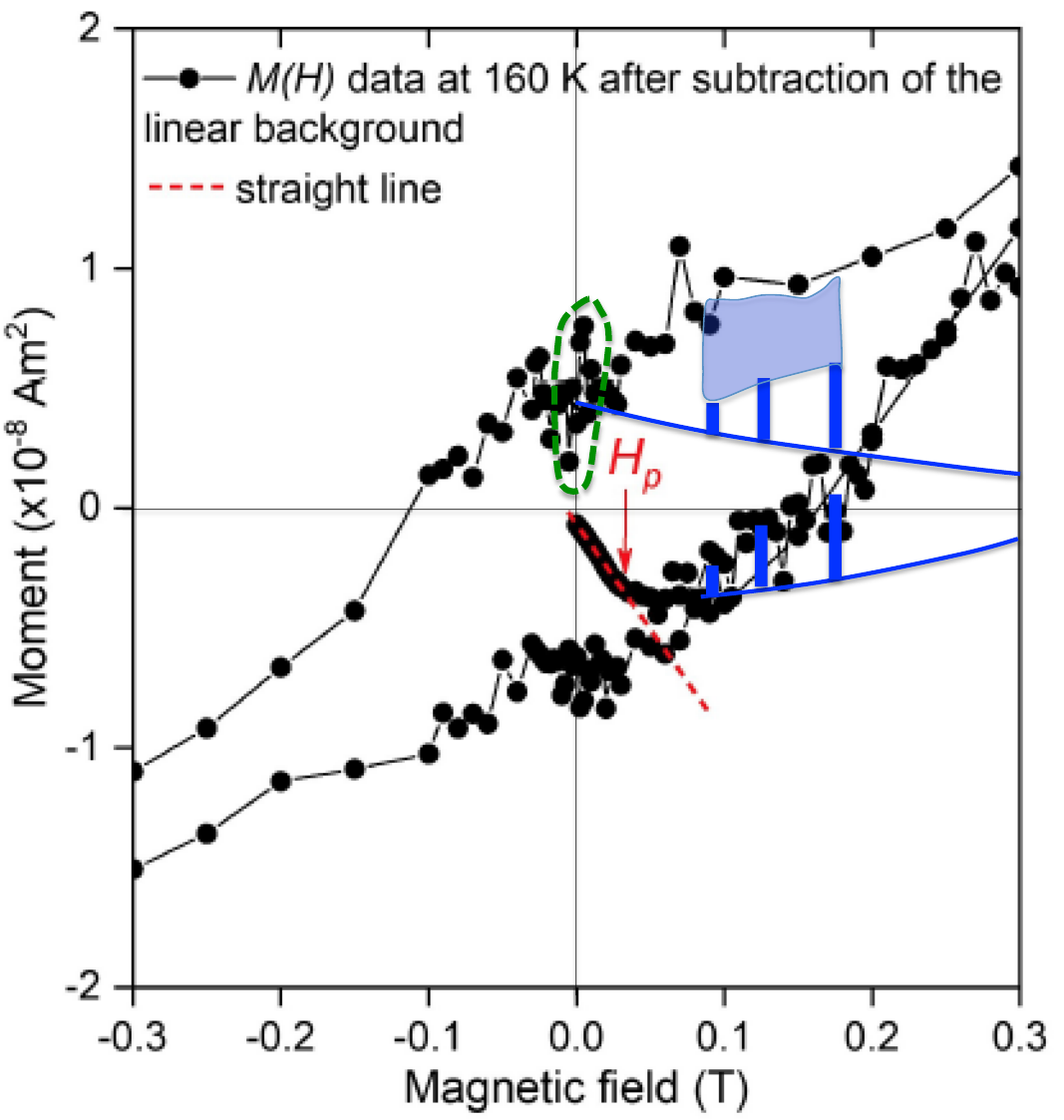}} 
 \caption { Fig. S12c of Ref. \cite{e2021p} showing magnetic moment versus field in a hysteresis cycle starting with the virgin curve.   We have  added the horizontal and vertical  lines going through the origin, the 
 dashed green  line showing the moment for zero field after half a cycle, and the blue lines and blue shaded area.
  Note that the moment  turns  positive for applied magnetic field 
  larger than 0.2 T. The blue additions are discussed in the text.
  }
  \label{figure1}
 \end{figure} 
 
            \begin{figure} [t]
 \resizebox{8.0cm}{!}{\includegraphics[width=6cm]{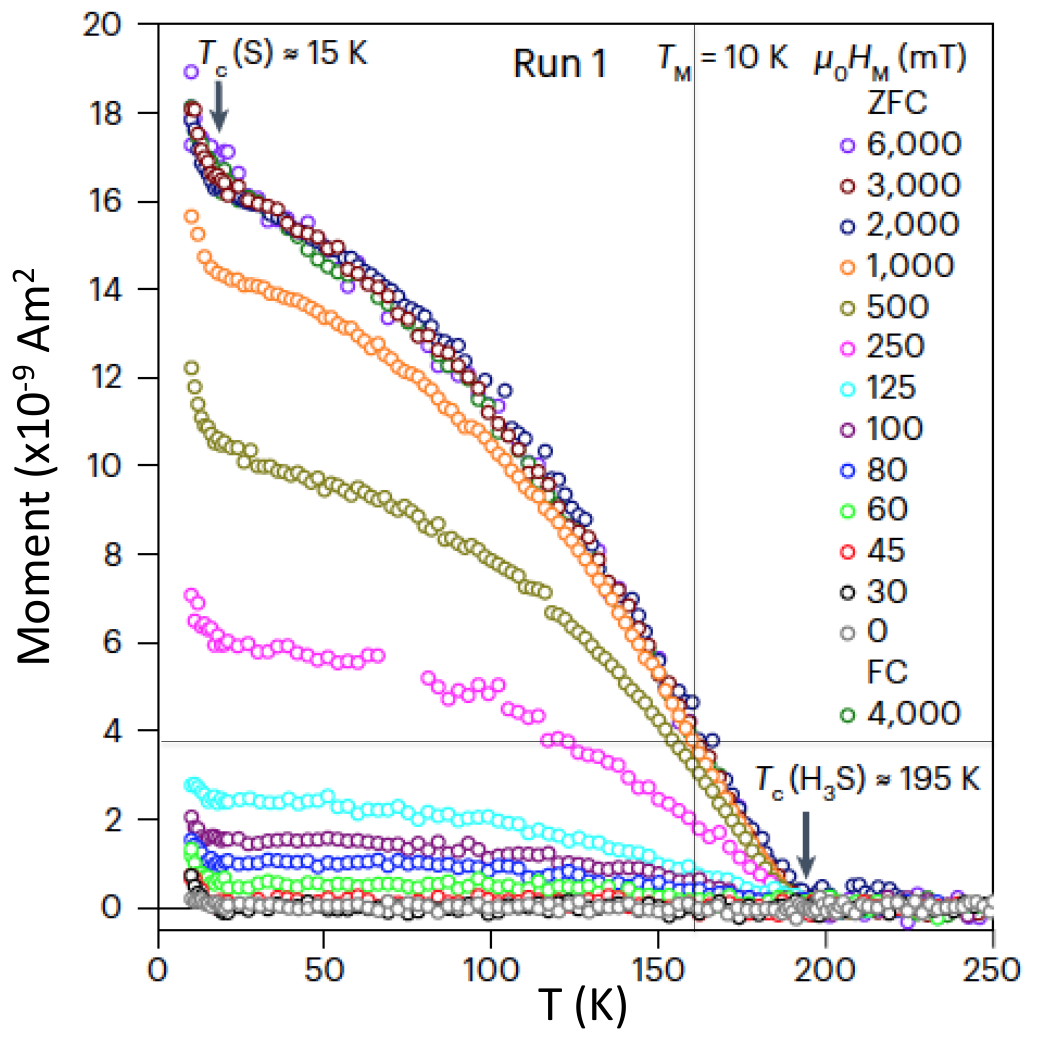}} 
 \caption { Fig. 1c of Ref. \cite{etrappedp}.    Measured magnetic moment under ZFC conditions. The field $H_M$ was applied at $T_M=10K$ and then
 removed. We added the vertical and horizontal thin lines indicating the trapped moment at $T=160K$ when the
 applied field was $H_M=1T$.
  }
  \label{figure1}
 \end{figure} 
 
           \begin{figure} [t]
 \resizebox{8.2cm}{!}{\includegraphics[width=6cm]{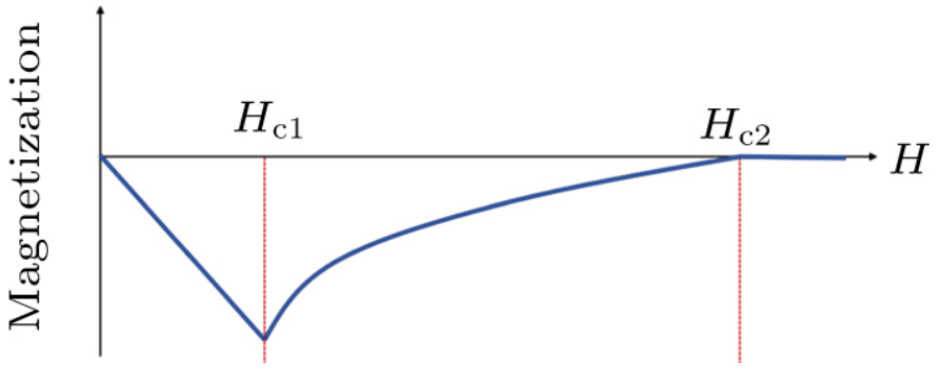}} 
 \caption {   Magnetization of an ideal type II superconductor as a function of applied magnetic field H.
 Note that the magnetization is negative up to applied field $H_{c2}$, the upper critical field.}
 \label{figure1}
 \end{figure}

%

Superconductors with strong pinning potentials will trap magnetic flux. In Ref.  \cite{etrappedp}, the magnetic moments that
were measured after an external magnetic field was applied and then removed were attributed to flux trapping due to
superconducting currents, which was interpreted as proof that the systems are superconductors.
Here we point out that the results reported in Ref. \cite{e2021p} by three of the authors of Ref. \cite{etrappedp}  invalidate the conclusions about
flux trapping of Ref. \cite{etrappedp}.

Fig. 1, taken from Fig. S12c of Ref. \cite{e2021p}, shows the magnetic moment versus magnetic field reported for $H_3S$ 
at 160K, after background subtraction \cite{correction}. The magnetic field was increased from zero to 1T and then decreased. The remnant magnetic moment
when the field reaches zero is, according to Fig. 1, approximately $m=0.4 \times 10^{-8} Am^2$.

Fig. 2, taken from Fig. 1c of Ref. \cite{etrappedp}, shows the measured magnetic moment versus temperature after a
magnetic field $H_M$ was applied at temperature $T_M=10K$ and then turned off, i.e. under ZFC conditions. Note that for applied
field $H_M=1T$, the trapped moment at $160K$ is approximately $m\sim 0.4 \times 10^{-8} Am^2$, consistent with
the result of Fig. 1. In Fig. 4a of Ref. \cite{etrappedp}, it is shown that for $T_M=100K$, the same result is  obtained for
the trapped moment at 160K, consistent with Fig. 1.

The value of the field where the magnetic moment starts to deviate from linearity in Fig. 1 was extracted to be $H_p(160K)=0.32mT$ \cite{e2021p}.
When corrected with a demagnetization factor estimated to be $1/(1-N)=8.5$, a lower critical field value was
obtained $H_{c1}(160K)=0.27 T$. The ratio $H_{c2}/H_{c1}$ for $H_3S$ was estimated to be 101 \cite{e2021p}. Therefore, at 160K the upper critical field should be
$H_{c2}(160K)=27 T$. When corrected for demagnetization, $H_{c2}(160K)/8.5=3.2T$.

A type II superconductor subject to an applied magnetic field will have a diamagnetic response, i.e. a negative magnetic moment,  for applied magnetic field up to the
upper critical field, as shown in Fig. 3. Therefore, if the magnetic moment depicted in Fig. 1 was due to a
superconducting sample,  the moment should remain negative after the field is first applied up to values of the field 27 T. Or, taking into 
account demagnetization, up to values of the field 3.2 T. Instead, Fig. 1 shows that the magnetic moment turns positive for
applied field larger than 0.2T.

This clearly indicates that the magnetic moment plotted in Fig. 1 cannot be interpreted as being the magnetic moment of a
 superconducting sample. Instead, it must reflect the magnetic response of the tiny sample together with the magnetic response of its 
environment and the measuring apparatus. If that is the case, there is no reason to believe that
the magnetic moment $m\sim 0.4 \times 10^{-8} Am^2$ for zero field seen in Fig. 1 reflects the magnetic moment of 
a superconducting sample, nor is there reason
to believe that the initial diamagnetic response shown in Fig. 1 is due to the sample and
not due to its environment.  Note that a large
diamagnetic background was subtracted from the measured results in Ref. \cite{e2021p, correction} to obtain the results shown in Fig. 1

The authors of Ref. \cite{etrappedp} would argue that even if the magnetic moment shown in Fig. 1 does show the response of 
a superconducting sample plus the background
at finite field,  there is no contribution from the background to the moment for zero applied field.
Indeed, Ref. \cite{etrappedp} states that {\it ``the signal of the trapped magnetic moment at zero field does not contain the field-dependent magnetic
background arising from the DAC''}. 
However, the only experimental evidence presented  in Ref. \cite{etrappedp} supporting that statement was that 
{\it ``No remnant nonlinear magnetic background
of the DAC body or other anomalies in the reference $m_{trap}(T)$ curve were
detected at ~10 GPa when the sample was evidently not superconducting''}. That of course says nothing about the behavior of the
background at pressure 155GPa at which the trapped flux measurements were conducted.
Note that the authors have not reported measurements at 155 GPa with non-superconducting samples showing no hysteresis.

Instead, we argue that Fig. 1 is incompatible with the assumptions made in Ref. \cite{etrappedp}  that (i) the sample is superconducting and
(ii) the magnetic moment at zero field is due to the sample. 
Assuming any plausible hysteretic loop for the sample consistent with the interpretation that it is
a hard type II superconductor, such as the blue thin lines shown in Fig. 1, the background that needs to be added to it to obtain the results shown in Fig. 1 would also be hysteretic.   The three lower thick blue segments in Fig. 1
show the distance between the hypothesized sample moment and the measured signal, which would be the background at those values of the
field as the field is increasing. Adding those thick blue segments to the hypothesized sample moment when the field is decreasing leaves
the blue shaded area as extra contribution from the background when the field is decreasing, that wasn't there when the field was
increasing. This would be clear evidence that the background is hysteretic.
A hysteretic background would indicate that there is ferromagnetism in the background, that   obviously would contribute or account for the entirety of   the
measured magnetic moment at zero field.

In conclusion, we argue that the magnetic moments measured in  Ref. \cite{etrappedp} shown in Fig. 2 and other figures of  Ref. \cite{etrappedp} cannot be interpreted 
to be the magnetic moment of  a superconducting sample as Ref. \cite{etrappedp} argues. One possibility is that they reflect the magnetic moment of a
ferromagnetic background, as discussed above. An alternative possibility is that the sample itself is ferromagnetic, and the initial
diamagnetic response in Fig. 1 is due to the background. Either of these possibilities would explain why
Ref. \cite{etrappedp} reported $linear$ dependence of ZFC moment versus field for small field rather than
quadratic as expected for a superconductor \cite{hmtrapped}.
To provide evidence that the hydride samples under pressure used in this study trap magnetic flux as hard superconductors do,
it is necessary to show that the samples show a hysteresis loop that is consistent with what superconducting samples of other known superconductors
 do, making the proper adjustments for differences in $H_{c1}$, $H_{c2}$, $T_c$ and strength and density of pinning centers.
Until that is done, it cannot be claimed that Ref. \cite{etrappedp} {\it ``proves
the existence of superconductivity in these materials''} as stated in Ref. \cite{etrappedp}.


\begin{references}
      
      \bibitem{etrappedp}     V. S.  Minkov et al,
``Magnetic flux trapping in hydrogen-rich high-temperature superconductors'',
\href{https://www.nature.com/articles/s41567-023-02089-1}{Nat. Phys. 19, 1293 (2023)}.


           \bibitem{e2021p} V. S. Minkov et al, ``Magnetic field screening in hydrogen-rich high-temperature superconductors'',
\href{https://www.nature.com/articles/s41467-022-30782-x} {Nat Commun 13, 3194 (2022)}.

\bibitem{correction} V. S. Minkov et al, ``Author Correction: Magnetic field screening in hydrogen-rich high-temperature superconductors'',
\href{https://www.nature.com/articles/s41467-023-40837-2}{Nat Commun 14, 5322 (2023)}.

\bibitem{hmtrapped} J. E. Hirsch and F. Marsiglio,
``Evidence Against Superconductivity in Flux Trapping Experiments on Hydrides Under High Pressure'', \href{https://link.springer.com/article/10.1007/s10948-022-06365-8}
{J Supercond Nov Magn 35, 3141 (2022)}.





                 \end{references}
 \end{document}